\documentclass[aps,prb,twocolumn]{revtex4}

\usepackage[dvips]{graphicx}

\renewcommand{\deg}{$^\circ$}

\begin{document}

\title{Gallium adsorption on (0001) GaN surfaces}

\author{Christoph Adelmann}
\altaffiliation{Present address: Department of Chemical Engineering and Materials Science, University of Minnesota, 151 Amundson Hall, 421 Washington Avenue SE, Minneapolis, Minnesota 55455}
\author{Julien Brault}
\author{Guido Mula}
\altaffiliation{Permanent address: INFM and Dipartimento di Fisica, Universit\`a di Cagliari, Cittadella Universitaria, Strada Provinciale Monserrato--Sestu km 0.700, 09042 Monserrato (CA), Italy}
\author{Bruno Daudin}
\affiliation{CEA/CNRS Research Group ``Nanophysique et Semiconducteurs'', D\'epartement de Recherche Fondamentale sur la Mati\`ere Condens\'ee, SPMM, CEA/Grenoble, 17 Rue des Martyrs, 38054 Grenoble Cedex 9, France}

\author{Liverios Lymperakis}
\author{J\"org Neugebauer}
\affiliation{Fritz-Haber-Institut der Max-Planck-Gesellschaft, \\Faradayweg 4-6, 14195 Berlin, Germany}

\begin{abstract}

We study the adsorption behavior of Ga on (0001) GaN surfaces combining experimental specular reflection high-energy electron diffraction with theoretical investigations in the framework of a
kinetic model for adsorption and {\em ab initio} calculations of energy parameters. Based on the experimental results we find that, for substrate temperatures and Ga fluxes typically used in
molecular-beam epitaxy of GaN, \emph{finite} equilibrium Ga surface coverages can be obtained. The measurement of a Ga/GaN adsorption isotherm allows the quantification of the equilibrium Ga surface coverage as a function of the impinging Ga flux. In particular, we show that a large range of Ga fluxes exists, where $2.5\pm 0.2$ monolayers (in terms of the GaN surface site density) of Ga are adsorbed on the GaN surface. We further demonstrate that the structure of this adsorbed Ga film is in good agreement with the laterally-contracted Ga bilayer model predicted to be most stable for strongly Ga-rich surfaces [J. E. Northrup \emph{et al.}, Phys. Rev. B \textbf{61}, 9932 (2000)]. For lower Ga fluxes, a discontinuous transition to Ga monolayer equilibrium coverage is found, followed by a continuous decrease towards zero coverage; for higher Ga fluxes, Ga droplet formation is found, similar to what has been observed during Ga-rich GaN growth. The boundary fluxes limiting the region of 2.5 monolayers equilibrium Ga adsorption have been measured as a function of the GaN substrate temperature giving rise to a Ga/GaN adsorption phase diagram. The temperature dependence is discussed within an {\em ab initio} based growth model for adsorption taking into account the nucleation of Ga clusters. This model consistently explains recent contradictory results of the activation energy describing the critical Ga flux for the onset of Ga droplet formation during Ga-rich GaN growth [B. Heying \emph{et al.}, J. Appl. Phys. \textbf{88}, 1855 (2000); C. Adelmann \emph{et al.}, J. Appl.
Phys. \textbf{91}, 9638 (2002).].

\end{abstract}
\pacs{68.43.-h;68.43.Bc;81.15.H;68.43.Mn}

\maketitle

\section{Introduction}

Due to its importance as a base material for the fabrication of optoelectronic devices in the blue and ultraviolet spectral region, GaN has been extensively studied in recent years. The comprehensive study of material and device properties\cite{Jain} has recently been complemented by an increasing number of studies concerning surface structures\cite{Hacke,Fritsch,Zywietz,Smith,Xue,Munkholm,Northrup1,Wang,Rapcewicz} and associated growth mechanisms, in particular for growth by plasma-assisted molecular-beam epitaxy (PAMBE) on the  Ga-polar (0001) surface.\cite{Tarsa,Widmann,Xie,Heying1,Bourret,Mula,Adelmann}

One of the fundamental results of these studies is that GaN growth by PAMBE ought to be carried out under Ga-rich conditions in order to obtain a smooth surface morphology and optimized material properties. At low growth temperatures, however, this leads to Ga accumulation and droplet formation, which is detrimental to the GaN epilayer quality.\cite{Heying2,Kruse} As a consequence, it was thought that optimum GaN growth conditions must be as close as possible to Ga/N stoichiometry. It has been recently observed that such Ga accumulation can be prevented when growing GaN at high temperatures and small Ga excess fluxes.\cite{Heying1,Adelmann} In particular, it has been shown that, at high growth temperatures, a wide range of Ga fluxes exists, for which a finite amount of
excess Ga is present on the GaN surface whose quantity is independent of the value of the Ga flux.\cite{Adelmann} Such conditions may provide a ``growth window'' for GaN PAMBE, i.e. a region, where the growth mechanisms and the surface morphology are independent of fluctuations of Ga flux and growth temperature.\cite{Heying1,Adelmann}

However, the quantitative description of a GaN ``growth diagram'', which describes the Ga surface coverage during growth as a function of Ga flux and growth temperature, has not yet been achieved: the results on the temperature dependence of the critical excess Ga flux at the onset of Ga droplet formation are contradictory, yielding activation energies of 2.8\,eV (Ref.~\onlinecite{Heying1}) and 4.8\,eV (Ref.~\onlinecite{Adelmann}), respectively. This suggests that the underlying mechanisms of Ga accumulation are not yet understood.

To address this discrepancy, we have performed Ga adsorption measurements on (0001) GaN. We discuss the results in the framework of a lattice-gas growth model for adsorption, which is based on {\it ab initio} calculated parameters. This model explains the origin of the apparently contradicting parameters derived from previous experimental studies\cite{Heying1,Adelmann} and gives a consistent description.

\section{Experimental Procedure}

The adsorption experiments were performed in a MECA2000 molecular-beam epitaxy chamber equipped with a standard effusion cell for Ga evaporation. The chamber also contains a rf plasma cell to provide active nitrogen for GaN growth. The pseudo-substrates used were 2\,$\mu$m thick (0001) (Ga-polarity)
GaN layers grown by MOCVD on sapphire. The substrate temperature $T_S$ was measured by a thermocouple in mechanical contact to the backside of the molybdenum sample holder and shielded from direct heating. Prior to all experiments, a 100 nm thick GaN layer was grown under Ga-rich conditions on the pseudosubstrates to remove the influence of a possible surface contamination layer.

Ga fluxes $\Phi$ have been calibrated to Ga effusion cell temperatures by reflection high-energy electron diffraction (RHEED) intensity oscillations during N-rich GaN growth at a substrate temperature of $T_S = 620$\,\deg C. In these conditions, the growth rate is actually proportional to the impinging Ga flux. It is sound to assume that the Ga adatom sticking coefficient is unity at such a low substrate temperature, which permits an absolute calibration of the Ga flux \emph{in terms of the GaN surface site density}.

The Ga surface coverage was assessed by analyzing the specularly-reflected RHEED intensity by a method described in Refs.~\onlinecite{Mula} and \onlinecite{Zheng}. This method uses the oscillatory transients in specular RHEED intensity, which are observed during Ga adsorption and desorption on/from (0001) GaN surfaces. It has been shown that the duration of these transients can be qualitatively related to the amount of adsorbing or desorbing Ga.\cite{Mula,Zheng}  In general, the relation between intensity and Ga coverage is unknown. Although tempting, the interpretation of these electron reflectivity transients in terms of RHEED oscillations is not obvious. Furthermore, it must been noted that the shape of the transients (albeit not their duration) depends on the diffraction conditions, notably the incidence angle. Of course, the modeling of electron reflection would allow
to directly relate the RHEED intensity to the Ga coverage (and the surface structure) but this is beyond the scope of this work.

However, the total duration of the transients occurring during Ga desorption can be used to qualitatively estimate the amount of Ga adsorbed on the surface. In Ref.~\onlinecite{Mula}, an indirect method has been used to draw limited \emph{quantitative} information from the desorption transients; below we will demonstrate a fully quantitative calibration relating the Ga desorption transient time to absolute Ga surface coverages. This allows us to circumvent the problem that the RHEED intensity cannot in general be easily related to adatom coverage. The experimental procedure is thus as follows: to assess the Ga quantity present after Ga adsorption for different impinging Ga fluxes, the Ga flux has been interrupted after a fixed adsorption time. The subsequent variation of the specular RHEED intensity due to Ga evaporation under vacuum has been recorded.

\section{Results}

\subsection{Ga adsorption}

\begin{figure}[tb]
\includegraphics[width=8cm,clip]{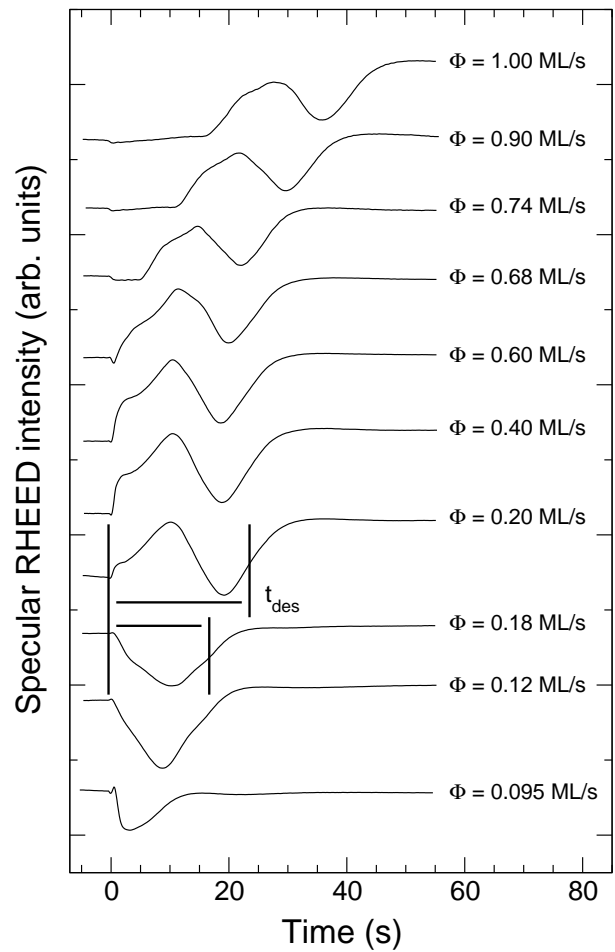}
\caption{\small \label{Fig:OR} Specular RHEED intensity during Ga desorption from a (0001) GaN surface. Beforehand, Ga adsorption has been carried out at Ga fluxes $\Phi$ as indicated. For $\Phi < 0.68$\,ML/s, the desorption transients correspond to equilibrium adsorption, i.e. they do not change as a function of the previous adsorption time. For $\Phi > 0.68$\,ML/s, this does not hold and the desorption transients depend on the adsorption time (here 1\,min). The substrate temperature is $T_S = 740$\,\deg C.}
\end{figure}

\begin{figure}[tb]
\includegraphics[width=8.5cm,clip]{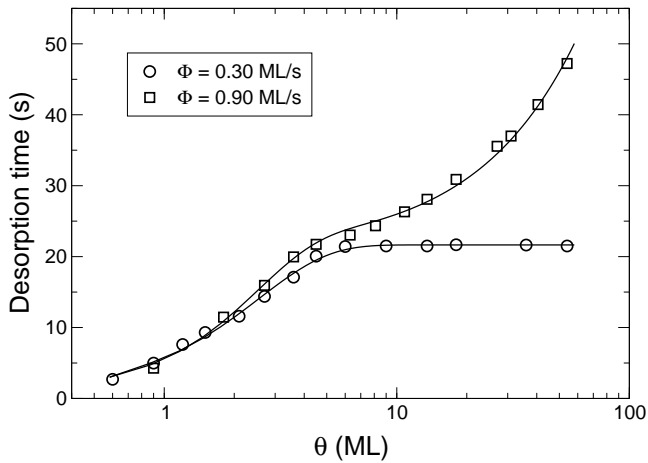}
\caption{\small \label{Fig:Sat} Ga desorption time as a function of the amount of nominally impinged Ga, $\theta$, for two different Ga fluxes, as indicated. $T_S = 740$\,\deg C.}
\end{figure}

Figure~\ref{Fig:OR} shows the variation of the specular RHEED intensity after 1\,min of Ga adsorption (substrate temperature $T_S = 740$\deg C, Ga fluxes $\Phi$ as indicated) and subsequent interruption of the Ga flux (at $t = 0$). We observe oscillatory transients during Ga desorption from the (0001) GaN surface. To quantify the desorption process, we define a desorption time $t_{\rm des}$ as the time interval between the shuttering of the Ga flux and the last inflection point in RHEED intensity
(Fig.~\ref{Fig:OR}). The desorption time depends on the amount of Ga present on the GaN surface after adsorption.\cite{Mula} The fundamental finding is that for Ga fluxes below $\Phi = 0.68$ monolayers (ML)/s, the desorption transients --- and thus the desorption time $t_{\rm des}$ --- are independent of the previous adsorption time, i.e. of the amount of nominally impinged Ga. This is consistent with the results in Ref.~\onlinecite{Mula} and is visualized in Fig.~\ref{Fig:Sat}, which shows $t_{\rm des}$ as a function of the amount of nominally impinged Ga $\theta$ (defined as the product of the Ga flux and the adsorption time). The derivative of this curve gives the (coverage and flux dependent) Ga sticking coefficient. Data are shown in Fig.~\ref{Fig:Sat} for a substrate temperature of $T_S = 740$\,\deg C and two different Ga fluxes: for $\Phi = 0.30$\,ML/s, the desorption time (and hence the amount of adsorbed Ga) monotonously increases until it saturates after $\theta \simeq 4$\,ML. At larger $\theta$, the adsorption has reached equilibrium and the coverage remains constant.

\begin{figure}[b]
\includegraphics[width=8.5cm,clip]{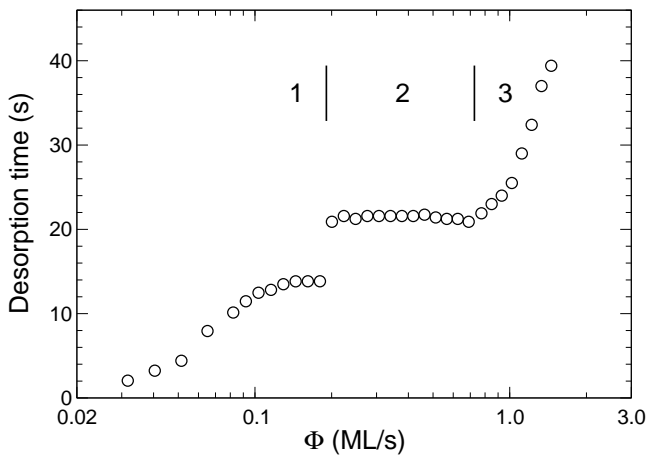}
\caption{\small \label{Fig:Isotherme} Ga desorption time as a function of impinging Ga flux $\Phi$ after equilibrium has been attained for regions 1 and 2, and after 1\,min of Ga adsorption in region 3. The substrate temperature is $T_S = 740$\,\deg C.}
\end{figure}

For a higher Ga flux of $\Phi = 0.90$\,ML/s, we observe the same behavior for $\theta \lesssim 4$\,ML, but thereafter, we find no saturation. Instead, a continuous increase of the desorption time
is observed, corresponding to Ga accumulation on the GaN surface. A more detailed analysis finds that the desorption transients for $\Phi \le 0.72$\,ML/s correspond to finite equilibrium Ga surface
coverages. For higher Ga fluxes, no finite equilibrium coverages exist and Ga will infinitely grow and finally form macroscopic droplets on the surface.

These results of Fig.~\ref{Fig:OR} are summarized in Fig.~\ref{Fig:Isotherme}, which shows the Ga desorption time --- related to the amount of adsorbed Ga --- as a function of the impinging Ga flux at a constant substrate temperature $T_S = 740$\deg C. It can be regarded as a Ga adsorption isotherm. We can discriminate three different regions:(1) an S-shaped increase of the Ga coverage for $\Phi < 0.20$\,ML/s, (2) a constant Ga coverage up to $\Phi = 0.72$\,ML/s, independent of the Ga flux, and (3) Ga accumulation and no finite equilibrium Ga coverages for higher $\Phi$. It is worth noting that the transitions between the three regimes are discontinuous within the experimental precision of 1\,\deg C of the Ga effusion cell ($\sim 7\times 10^{-3}$\,ML/s for fluxes around 0.5\,ML/s). In particular, no intermediate equilibrium coverages have been found between regimes 1 and 2. Therefore, the transition fluxes between the different regimes are well defined.

\begin{figure}[tb]
\includegraphics[width=8.5cm,clip]{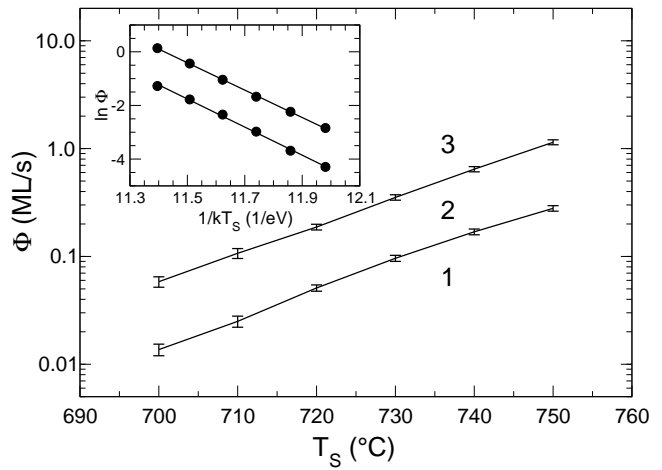}
\caption{\small \label{Fig:DDP} Ga adsorption phase diagram indicating the Ga surface coverage as a function of impinging Ga flux $\Phi$ and substrate temperature $T_S$. The definition of regions 1--3 follows Fig.~\ref{Fig:Isotherme}. The inset shows an Arrhenius plot of the data.}
\end{figure}

\begin{table*}[tb]
\begin{ruledtabular}

\caption{\small Experimental activation energies $E_{A}^{\rm{exp}}$ and prefactors $\nu_{\rm{des}}^{\rm{exp}}$ for the transition fluxes between different Ga coverage regimes as obtained from the adsorption (this work, Fig.~\ref{Fig:DDP}) and growth phase diagrams (from Refs. as indicated). $v$ denotes the GaN growth rate (the N flux), $\alpha$ the linear temperature coefficient of the adsorption energy, and $\nu_{\rm{des}}^{\rm{ren}}$ the renormalized prefactor (see Sec.~IV). \label{Tab:FitResults}}

\begin{tabular}{llccccc}

Reference & Transition & $E_{A}^{\rm{exp}}$ (eV) &
$\nu_{\rm{des}}^{\rm{exp}} (\text{Hz})$ & $v$ (ML/s) & $\alpha$
({\rm meV/K}) & $\nu_{\rm{des}}^{\rm{ren}} (\text{Hz})$\\ \hline

This work & $1\rightarrow 2$ & 5.2 &  $3\times 10^{25}$ & 0  & $-2.3$  &   $4.7\times10^{25}$ \\

This work & $2\rightarrow 3$ & 5.1 & $2\times 10^{25}$ & 0  & $-2.2$  &   $1.5\times 10^{25}$ \\

Ref.~\onlinecite{Adelmann} &$1\rightarrow 2$ & 3.7 & $5\times 10^{17}$ & 0.28  & $-0.8$ & $1.6\times 10^{18}$ \\

Ref.~\onlinecite{Adelmann} & $2\rightarrow 3$ & 4.8 &  $1\times 10^{24}$ & 0.28  & $-1.9$ & $4.8\times 10^{23}$ \\

Ref.~\onlinecite{Heying1} & $2\rightarrow 3$ & 2.8 & $1 \times 10^{14}$ & 1.1  &  $> - 0.1$ & $5.5\times 10^{13}$ \\

\end{tabular}

\end{ruledtabular}
\end{table*}

To fully assess the thermodynamics of the adsorption process, the variation of the adsorption isotherm has to be known as a function of substrate temperature. In the following we will restrict
ourselves to the study of the variation of the transition fluxes between the different regimes. The result is plotted in Fig.~\ref{Fig:DDP}. We see that the transition fluxes vary exponentially with substrate temperature. The inset shows an Arrhenius plot of the data, yielding activation energies of
$E_A^{(12)} = 5.2 \pm 0.1$\,eV and $E_A^{(23)} = 5.1 \pm 0.05$\,eV for the $1\rightarrow 2$ and $2\rightarrow 3$ transition, respectively. These values, the prefactors, and the corresponding values for the growth phase diagram in Refs.~\onlinecite{Adelmann} ($1\rightarrow 2$ corresponds to $B\rightarrow C$ and $2\rightarrow 3$ corresponds to $C\rightarrow D$) and \onlinecite{Heying1} (discussing the $2\rightarrow 3$ transition only) are summarized in Tab.~\ref{Tab:FitResults} and will be discussed in Sec.~IV.

\subsection{GaN surface structures}

The adsorption isotherm in Fig.~\ref{Fig:Isotherme} is given in terms of the Ga desorption time, which is only \emph{qualitatively} related to the amount of adsorbed Ga. For a fully \emph{quantitative} treatment of Ga adsorption, one must know the dependence of the Ga desorption rate $\Phi_{\rm des}$ on the Ga surface coverage $c$. This would allow the modeling of the Ga re-evaporation, as the desorption rate is given by

\begin{equation}
\label{Eq:RateEq} \Phi_{\rm des}(c) = \frac{{\rm d}c}{{\rm d}t}
\quad ,
\end{equation}

\noindent with the initial condition $c(t=-t_{\rm des}) = c_{\rm eq}$, which denotes the amount of Ga adsorbed in equilibrium conditions, i.e., before Ga desorption sets in. After the time interval $t_{\rm des}$, the Ga coverage becomes zero. The knowledge of $t_{\rm des}$ would thus allow us to compute $c_{\rm eq}$ if $\Phi_{\rm des}(c)$ was known (which it is not). This requirement can be circumvented by considering that in equilibrium, the impinging Ga flux $\Phi$ must exactly balance the evaporation rate, hence $\Phi_{\rm des}(c_{\rm eq}) = - \Phi(c_{\rm eq}[t_{\rm des}])$. Integrating Eq.~(\ref{Eq:RateEq}), taking the first derivative with respect to $t_{\rm des}$, and
using the above substitution leads to 

\begin{equation}
\label{Eq:ModelAdsorptionIsotherm}
c_{\rm eq}(\Phi) = \int_0^{\Phi} \Phi'\frac{\partial t_{\rm des}}{\partial\Phi'} {\rm d}\Phi'\quad .
\end{equation}

\noindent This expression allows the computation of $c_{\rm eq}$ from $t_{\rm des}$ as a function of $\Phi$ (which is known from the experimental data) and can be evaluated numerically. Note that
$c_{\rm eq}$ only depends on the derivative of $t_{\rm des}(\Phi)$, which means that $t_{\rm des}$ does not necessarily have to denote the very end of Ga adsorption but can be taken after any time interval, as long as it is well-defined in the RHEED signal. Any further Ga desorption after the end of the chosen time interval will lead to a constant offset $t_{\rm des}$, which does not contribute to $c_{\rm eq}$ in Eq.~(\ref{Eq:ModelAdsorptionIsotherm}).

\begin{figure}[b]
\includegraphics[width=8.5cm,clip]{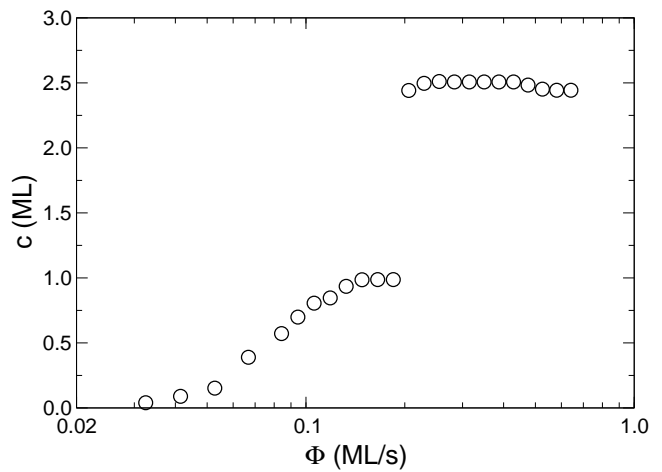}
\caption{\small \label{Fig:Calib} Calibrated Ga adsorption isotherm at $T_S = 740$\,\deg C using Eq.~(\ref{Eq:ModelAdsorptionIsotherm}). The data are derived from Fig.~\ref{Fig:Isotherme}. As the method can only be applied to equilibrium coverages, only regions 1 and 2 are represented.}
\end{figure}

Using Eq.~(\ref{Eq:ModelAdsorptionIsotherm}), we can now calibrate the Ga adsorption isotherm in Fig.~\ref{Fig:Isotherme}. Since the relation $\Phi_{\rm des}(c_{\rm eq}) = - \Phi(t_{\rm des})$ implies steady-state conditions, only the data in regions 1 and 2 can be treated. Applying Eq.~(\ref{Eq:ModelAdsorptionIsotherm}) to the data in region 3 does not lead to the amount of Ga adsorbed after a finite time but to incorrect values because under these conditions $\vert\Phi_{\rm des}\vert < \vert\Phi\vert$. Since experimental data are available only for $\Phi > 0.032$\,ML/s, the isotherm has been extrapolated by an exponential for smaller fluxes. The contribution of the interval $0 < \Phi < 0.032$\,ML/s to the overall integral in Eq.~(\ref{Eq:ModelAdsorptionIsotherm}) is 0.04\,ML. This is only a minor correction, suggesting that the specific form of the extrapolation function is not important. The result of the calibration is plotted in Fig.~\ref{Fig:Calib}. We observe that in region 1, the Ga coverage increases from almost zero to a value of 0.98\,ML, close to 1\,ML. The coverage then increases abruptly to a value of 2.5\,ML in region 2. Typical systematical errors can be estimated to be of the order of $\pm 0.2$\,ML.

What is the detailed structure of such adsorbed Ga films? It must be kept in mind that the Ga fluxes have been calibrated by GaN RHEED oscillations and are hence given in terms of the GaN surface site density. A GaN coverage of 1\,ML thus indicates the adsorption of a Ga adatom on each GaN site. This suggests that Ga adsorbs in region 1 as a coherent (pseudomorphic) adlayer. 

At the $1\rightarrow 2$ transition, the Ga coverage increases by about 1.5\,ML, i.e. by more than a pseudomorphic adlayer. This compares favorably to the laterally-contracted bilayer model, which has been calculated to be the most stable structure of Ga-rich (0001) GaN surfaces.\cite{Northrup1} It consists of two adsorbed Ga adlayers on top of a Ga-terminated GaN surface. The first layer is found to be pseudomorphic to the GaN surface but the second one has an in-plane lattice parameter of 2.75\,\AA, about 13.8\% smaller than that of GaN (3.189\,\AA{}). The second layer thus contains 1.3\,ML of Ga in terms of the GaN surface site density, in good agreement with the experimental value of 1.5\,ML.

\begin{figure}[tb]
\includegraphics[width=8cm,clip]{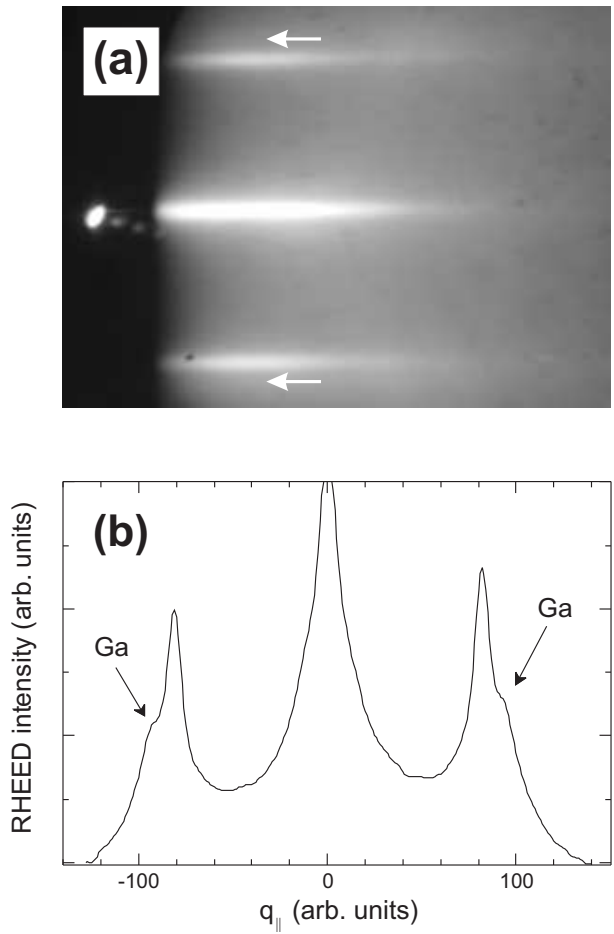}
\caption{\small \label{Fig:RHEED} \textbf{(a)} RHEED pattern (azimuth $\langle 11\bar{2}0 \rangle$) of a (0001) GaN surface with a Ga bilayer present at $T_S = 600$\,\deg C. The white arrows indicate additional streaks due to the Ga film. \textbf{(b)} RHEED intensity profile in the $\langle 10\bar{1}0 \rangle$ direction evidencing the additional RHEED streaks due to the Ga bilayer.}
\end{figure}

The formation of a laterally-contracted bilayer structure in regime 2 is further corroborated by the observation of supplementary streaks in the RHEED pattern after Ga adsorption in regime 2 and rapid quenching of the sample down to substrate temperatures below about $T_S = 650$\,\deg C [see Fig.~\ref{Fig:RHEED}(a)]. At higher temperatures, the supplementary streaks are too weak to be detected unambiguously. The lattice parameter corresponding to these streaks is found to be $2.73 \pm 0.03$\,\AA{} by a fit using pseudo-Voigt functions and assuming the bulk lattice parameter for the GaN layer. This value compares very favorably to that obtained by the \emph{ab inito} calculations (2.75\,\AA{}).\cite{Northrup1} Combined with the adsorption results, this strongly suggests that adsorption in region 2 leads to the formation of a laterally-contracted Ga bilayer.

The shape of the oscillation transients in Fig.~\ref{Fig:OR} suggests that, in region 3, Ga droplets are formed on top of this Ga bilayer. This Ga de-wetting transition may indicate that the attractive interaction energy of a Ga adatom with the surface is maximum in the second layer and lower in the third layer.\cite{Brochard} This conclusion is consistent with the first principles results which will be discussed in the next section. Ga thus grows in a Stranski-Krastanow mode on (0001) GaN. 

Finally, the different adsorption regimes can be summarized as follows: (1) Ga coverage $c \le 1$\,ML, i.e. successive formation of a coherent Ga monolayer, (2) a Ga coverage of $c = 2.5$\,ML, forming a laterally-contracted Ga bilayer, and (3) Ga accumulation and droplet formation on top of a Ga bilayer. In the following, we will derive an ab initio based growth model which describes the temperature dependence of the transition fluxes between the different regimes.

\section{Discussion}

\begin{figure}[tb]
\includegraphics[width=8cm,clip]{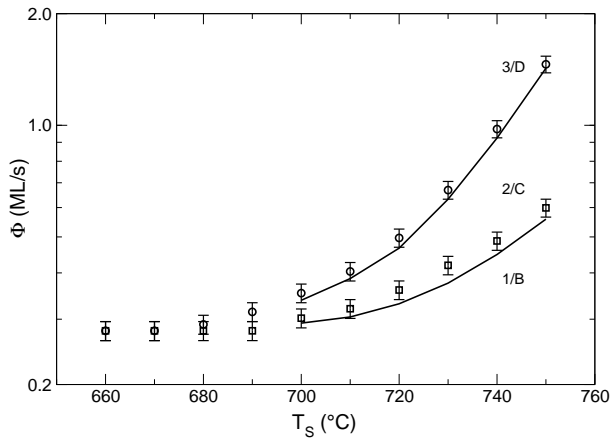}
\caption{\small \label{Fig:Comp} Comparison between the experimental growth phase diagram in Ref.~\onlinecite{Adelmann} (open symbols) and the diagram derived from the the Ga adsorption data in this work (solid lines). The different regimes 1, 2, and 3 of the adsorption phase diagram are indicated, which correspond to regimes B, C, and D in Ref.~\onlinecite{Adelmann}, respectively.}
\end{figure}

An intuitive connection between adsorption and growth phase diagrams can be made by assuming that the \emph{excess} Ga in Ga-rich growth conditions behaves as if it would be adsorbed on a GaN surface. Then, adding the GaN growth rate (0.28\,ML/s in the experiments of Ref.~\onlinecite{Adelmann}) to the adsorption phase transition fluxes should reproduce the growth phase diagram of Ref.~\onlinecite{Adelmann}. This comparison is shown in Fig.~\ref{Fig:Comp}. It demonstrates good overall agreement, suggesting that the assumption is valid. However, it must be noted that the critical fluxes derived from adsorption measurements fall below those in the growth phase diagram, although differences are of the order of the experimental precision.

Yet, a closer look at the activation barriers and prefactors (Tab.~\ref{Tab:FitResults}), as derived from the experimental phase diagrams, poses a number of questions. First, why are the activation energies and prefactors in the adsorption and growth phase diagram so different, even though absolute values of transition fluxes are close? Second, why are the values for the same transition ($2\rightarrow 3$ for growth) in Refs.~\onlinecite{Heying1} and \onlinecite{Adelmann} so different? Finally, while the barrier of 2.8\,eV for the $2\rightarrow 3$ transition in the growth phase diagram in Ref.~\onlinecite{Heying1} is close to the cohesive energy of bulk Ga and has thus been directly interpreted as a Ga desorption barrier, what is the physical origin of the 5.1\,eV activation energy and the meaning of a prefactor of 10$^{25}$\,Hz? The latter value is fundamentally
different from prefactors typically observed/calculated for diffusion or desorption processes (which are of the order of $\sim 10^{13}$\,Hz).

\subsection{Growth and adsorption model}

To address the above questions, we have analyzed the data in terms of a simple growth model and in combination with first principles total energy calculations. In order to simplify the discussion of the growth model, we divide the problem in two parts: First, we derive how the density of Ga adatoms $\rho$ on the surface\cite{footnote} depends on parameters such as temperature $T$, Ga flux $\Phi$, and growth rate $v$. Second, we calculate the critical adatom density, at which nucleation occurs and the system undergoes a phase transition.

The adatom coverage is given by:

\begin{equation}\label{eq:drhodta}
  \frac{\rm{d} \rho (\mathbf{r},t)}{\rm{d}t} = D \nabla^2
  \rho(\mathbf{r},t)+\Phi-\frac{1}{\tau_{\rm{inc}}}\rho
  (\mathbf{r},t)-\frac{1}{\tau_{\rm{des}}}\rho
  (\mathbf{r},t) \quad .
\end{equation}

\noindent Here, the desorption time $\tau_{\rm des}$ is given by

\begin{equation}\label{eq:des}
  \tau_{\rm des}^{-1} = \nu_{\rm des} e^{-E_{\rm des}/k_B T}\quad ,
\end{equation}

\noindent with $E_{\rm des}$ the desorption barrier and $k_B$ the Boltzmann constant. $D$ is the surface diffusion constant for Ga adatoms and $\mathbf{r}$ gives the lateral position on the surface. For step flow growth (as realized in Ga-rich growth conditions\cite{Adelmann}), incorporation occurs essentially at the step edges. Since these are moving, the incorporation rate ${\tau^{-1}_{\rm inc}(\mathbf{r},t)}$ is in general inhomogeneous and time dependent. For ideal step flow it will be zero on the terraces and $>0$ only at the step edges. The incorporation rate and the growth rate $v$ are directly related by:

\begin{equation}
  v(t) = \frac{1}{A} \int \frac{1}{\tau_{\rm inc}(\mathbf{r},t)}\rho(\mathbf{r},t) d\mathbf{r} \quad .
\end{equation}

\noindent Here the integration is performed over the total surface area $A$. Since in our experimental setup stationary conditions are realized, the explicit time dependence in the growth rate disappears (i.e. $v(t)=v$). Furthermore, in stationary conditions Eq.~(\ref{eq:drhodta}) also simplifies: $\mathrm{d}\rho/\mathrm{d}t=0$, i.e. the left hand side becomes zero. A further simplification in Eq.~(\ref{eq:drhodta}) can be made by taking into account that the phase transitions we are interested in occur exclusively under Ga-rich conditions. Under these conditions the surface steps on the surface are Ga-terminated\cite{StepsToBePublished} and the incorporation rate $\tau^{-1}_{\rm {inc}}$ at such a step will be small. This is in contrast to the conventional step flow picture where the sticking probability of an atom at the step edge is assumed to be close to one. The difference to the conventional model is due to the fact that we have two species with very different concentrations. For nitrogen (which is the minority species for very Ga-rich conditions) the sticking coefficient at the Ga-terminated step edges will be close to one and steps act as sinks to the nitrogen concentration which becomes highly inhomegenous. For Ga atoms, however, the sticking probablity is low. Thus, the effect of steps on Ga will be small and the Ga-adatom density is virtually homogeneous, i.e. $D\nabla^2 \rho(\mathbf{r}) = 0$.

Based on the above discussion, Eq.~(\ref{eq:drhodta}) can be written as:

\begin{equation}\label{eq:drhodt}
  0 = \Phi - v - \frac{1}{\tau_{\rm des}}\rho_0 \quad .
\end{equation}

\noindent Here, $\rho_0$ denotes the equilibrium adatom density. We note that this equation holds for both growth ($v>0$) and adsorption ($v=0$). The solution is easily found to be

\begin{equation}\label{eq:rho}
   \rho_0 = \left({\Phi - v}\right){\tau_{\rm des}} \quad .
\end{equation}

Nucleation occurs if the (stable) nuclei are in thermodynamic equilibrium with the lattice gas (which is described by the adatom density) on the surface. At low densities ($\rho_0\ll 1$), interactions in the lattice gas itself can be neglected and we obtain

\begin{equation}\label{eq:rhocrit}
   \frac{N_{\rm ad}}{N_{\rm sites}} = \rho_{\rm crit} = e^{-\Delta E_{\rm
nuc}/k_B T} \quad .
\end{equation}

\noindent Here, $N_{\rm ad}$ is the number of adatoms in the lattice gas, $N_{\rm sites}$ is the total number of surface sites which can be occupied by the adatoms, and $\Delta E_{\rm nuc}$ is
the energy the adatom gains if it is attached to a subcritical nucleus making the latter stable.

Based on the above model, we can directly obtain the critical Ga flux $\Phi_{\rm crit}$ at  which the phase transitions occur. Combining Eqs.~(\ref{eq:rho}) and (\ref{eq:rhocrit}) gives:

\begin{equation}\label{eq:critFlux}
\Phi_{\rm crit} = v + \tau_{\rm des}^{-1} e^{-\Delta E_{\rm nuc}/k_B T} \quad .
\end{equation}

\noindent This equation can be rewritten using Eq.~(\ref{eq:des}) as

\begin{equation}\label{eq:critFlux2}
  \Phi_{\rm crit}= v + \nu_{\rm des} e^{-(E_{\rm des}+\Delta E_{\rm nuc})/k_B T}
\end{equation}

\noindent and applies both to the adsorption and growth phase diagram. The activation energy is thus expected to represent the total binding energy of a Ga atom in a critical Ga cluster.

\subsection{First principles analysis}

A comparison with the experimental results (see Tab.~\ref{Tab:FitResults}) shows that energy and prefactor are {\em not} constant [as expected from Eq.~(\ref{eq:critFlux2})] but vary largely with the growth conditions. As a general trend, one finds that the activation energy and the prefactor decrease with increasing growth rate (N flux). It is also interesting to note that only in the case of high growth rate ($v = 1.1$\,ML/s) the prefactor ($1\times 10^{14}$\,Hz) is close to the typical attempt frequencies observed/expected for desorption, i.e. in the $10^{13}$\,Hz range. For conditions where growth is slow ($v = 0.28$\,ML/s) or absent (adsorption), prefactors are found which are many orders of magnitude larger.

In order to identify the origin of these apparent discrepancies, we have explicitly calculated the desorption and the formation of small Ga clusters on the Ga bilayer surface employing density functional theory. In the following we will focus on the Ga-bilayer structure (which corresponds to the $2\rightarrow 3$ transition). Based on the almost identical energies of the $1 \rightarrow 2$ and $2\rightarrow 3$ transitions in the adsorption phase diagram we expect the mechanisms/energetics to be rather similar.

\begin{figure}[tb]
\includegraphics[width=8cm,clip]{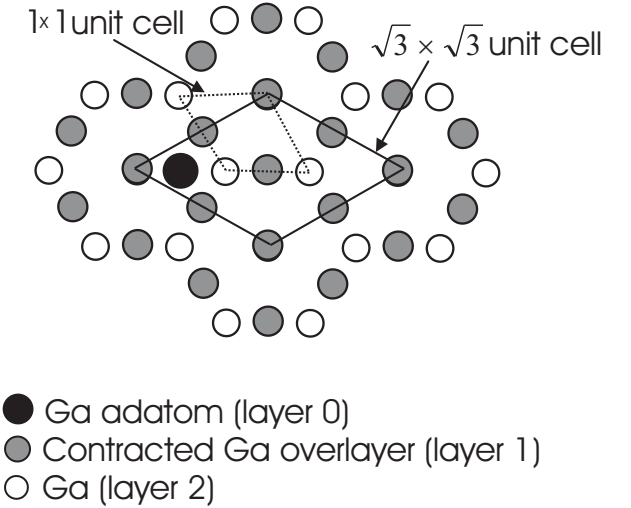}
\caption{\small \label{Fig:Schema1} Schematic top view of an adatom on the Ga bilayer structure. The white balls mark the positions of the Ga atoms in the second layer. The gray balls mark the positions of the Ga atoms in the contracted Ga epilayer and the black ball the Ga adatom in the T4 position. The dashed line shows the $1\times 1$ surface unit cell of the ideal bulk truncated GaN
(0001) surface. The solid line shows the surface unit cell of the Ga bilayer structure.}
\end{figure}

\begin{figure}[b]
\includegraphics[width=8cm,clip]{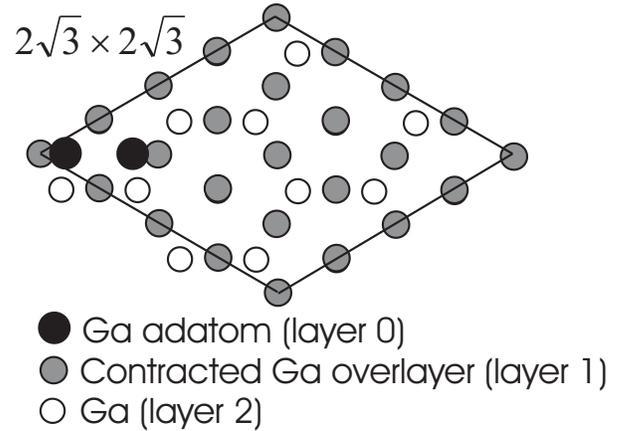}
\caption{\small \label{Fig:Schema2} Schematic top view of a 2 atom island (dimer) on the Ga bilayer structure. The solid line shows the $\left( 2\sqrt3\times2\sqrt3\right)$ surface unit cell which has been used to describe this structure.}
\end{figure}

Specifically, we use soft Troullier-Martins pseudopotentials\cite{Troullier} and the Perdew-Burke-Ernzerhof generalized-gradient approximation (PBE-GGA) to describe exchange/correlation.\cite{PBE} The Ga $3d$ semicore states were described in the frozen core approximation (nlcc).\cite{Loui82} Details about the method can be found elsewhere.\cite{Bockstedtexx} The calculations have been performed with a plane wave basis set (energy cutoff: 50\,Ry). The Brillouin zone has been sampled by a $2\times 2 \times 1$ Monkhorst-Pack mesh for the $\left( 2\sqrt 3 \times 2\sqrt 3\right)$ unit cell and $4\times 4 \times 1$ for the $\left( \sqrt 3 \times \sqrt 3\right)$ unit cell.\cite{Monkhorstxx} The Ga bilayer with one up to four adatoms has been modeled by a slab consisting of 2 double layers and $\left(\sqrt 3 \times \sqrt 3\right)$ (to describe the adsorption of a single adatom, see Fig.~\ref{Fig:Schema1}) and larger $\left( 2\sqrt 3 \times 2\sqrt 3\right)$ (to describe adatom islands, see Fig.~\ref{Fig:Schema2}) unit cells. Increasing the slab thickness to 4 double layers changes the surface energy by less than 0.5\,meV per surface unit cell. Adatoms are added only on the upper surface of the slab. The lower surface has been passivated by pseudo-hydrogen to remove the electrically active surface states. The adatom(s) and the first two surface layers have been fully relaxed. Detailed convergence checks can be found in Refs.~\onlinecite{mrsProc,mPlane,Fuchs}.

Based on these studies, we have calculated the desorption energy of an adatom and the binding energy of adatoms in small islands. The desorption energy (the energy needed to remove the adatom from the surface) is defined by 

\begin{equation}
   E_{\rm des} = -\left( E_{\rm tot}^{\rm adatom} - E_{\rm tot}^{\rm slab} - E_{\rm
tot}^{\rm atom}\right)\quad ,
\end{equation}

\noindent where $E_{\rm tot}^{\rm adatom}$ is the total energy of the surface including the adatom, $E_{\rm tot}^{\rm slab}$ that of the free surface and $E_{\rm tot}^{\rm atom}$ that of the (spin-polarized\cite{spinPolarization}) Ga atom. Using this expression we find an adatom binding energy of 2.52\,eV for the $\left(\sqrt 3 \times \sqrt 3\right)$ structure and 2.41\,eV for the $\left(2\sqrt 3 \times 2\sqrt 3\right)$ unit cell. In the equilibrium configuration, the adatom sits on a three-fold coordinated hollow site. The Ga--Ga bond length between adatom and surface layer is 2.68\,\AA{}, i.e. very close to the nearest neighbor distance in $\alpha$-Ga of 2.71\,\AA{}.\cite{Wyckoff} 

The binding energy of an adatom in an island consisting of $n_{\rm ad}$ adatoms is given by:

\begin{equation}
   E_{\rm isl} = -\left[\frac{1}{n_{\rm ad}}\left( E_{\rm tot}^{\rm isl} - E_{\rm tot}^{\rm
slab} - E_{\rm tot}^{\rm atom}\right) - E_{\rm des}\right] \quad .
\end{equation}

\noindent For an island consisting of 2 adatoms we find $E_{\rm isl}=0.15$\,eV. For larger islands consisting of 3 adatoms $E_{\rm isl}=0.30$\,eV and $E_{\rm isl}=0.34$\,eV for a 4 atom island. The
numbers are the island formation energies as calculated in a $\left( 2\sqrt3 \times 2\sqrt3\right)$ cell. It is interesting to note here that all islands are unstable against the formation of Ga-droplets: the formation energy of a Ga atom in an island ($E_{\rm des} + E_{\rm isl} \simeq 2.75$\,eV) is smaller than the cohesive energy of bulk Ga of 2.8\,eV.\cite{Kittel} Therefore the islands act as nucleation centers for Ga droplet formation.

\subsection{Interpetation of the results}

\begin{figure}[tb]
\includegraphics[width=8cm,clip]{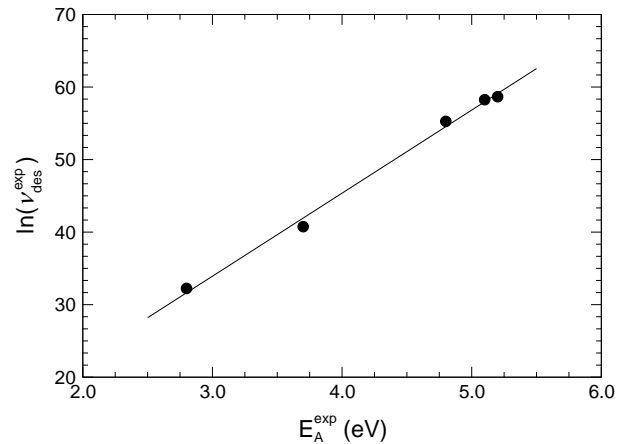}
\caption{\small \label{Fig:lnnuEA} Natural logarithm of the experimental prefactors $\ln\left(\nu_{\rm{des}}^{\rm{exp}}\right)$ as a function of the activation energy $E_A^{\rm{exp}}$ for various data in this work and Refs.~\onlinecite{Heying1} and \onlinecite{Adelmann} (full circles; see Tab.~\ref{Tab:FitResults}). The solid line represents a linear fit.}
\end{figure}

We can now compare these energies with those obtained from the analysis of the experimental data (Tab.~\ref{Tab:FitResults}). As can be seen, the activation energy is close to our calculated desorption energy only for the high growth case (1.1\,ML/s). For lower growth rates or adsorption, activation energies are found which are way too large. A closer analysis of the prefactors and the activation energies shows a clear relation (see Fig.~\ref{Fig:lnnuEA}): 

\begin{equation}\label{eq:fit}
   \ln\left(\nu_{\rm{des}}^{\rm{exp}}\right) = a_1 E_{\rm{A}}^{\rm{exp}} + a_0 \quad .
\end{equation}

\noindent The supercript "exp" indicates experimental data. For the values given in Tab.~\ref{Tab:FitResults} we get $a_1 = 11.45$\,eV$^{-1}$ and $a_0 = -0.427$. 

The observed relation between prefactor and activation barrier can be explained in terms of a temperature dependent activation energy. Since the experimentally accessible temperature range is
rather small ($\simeq 50$\,K), we assume a linear dependence: $E_A = E_0 + \alpha (T-T_0)$ with $E_0$ the temperature independent contribution, $T_0$ the temperature offset, and $\alpha$ the linear temperature coefficient. Equation~(\ref{eq:critFlux2}) then becomes

\begin{equation}\label{eq:critFluxlinEA}
  \Phi'_{\rm crit} = v + \nu_{\rm des}e^{-\alpha/k_B} e^{-(E_0-\alpha T_0)/k_B
T} \quad .
\end{equation}

\noindent A comparison between  Eqs.~(\ref{eq:fit}) and (\ref{eq:critFluxlinEA}) gives the following relations:

\begin{equation}\label{eq:expEa}
E_{\rm{A}}^{\rm{exp}} = E_0 - \alpha T_0 \qquad \text{and} \qquad
\ln\left(\nu_{\rm{des}}^{\rm{exp}}\right) = \ln\left(\nu_{\rm des}\right) -
\frac{\alpha}{k_B} \quad .
\end{equation}

\noindent This leads to

\begin{eqnarray}
   a_0 = \ln(\nu_{\rm des}) - \frac{E_0}{k_BT_0} \qquad \text{and} \qquad a_1 =
\frac{1}{k_B T_0} \quad .
\end{eqnarray}

Using these relations and $E_0=2.75$\,eV (as found from our first principles calculations), we obtain $T_0 = 740$\,\deg C (which is close to the experimentally accessed temperature range, so the linear approximation for the temperature dependence of $E_A$ is well justified) and $\nu_{\rm des} =  3\times 10^{13}$\,Hz (which is close to typical attempt frequencies).

Using the above parameters and Eq.~(\ref{eq:expEa}), the linear temperature coefficient $\alpha = (E_0-E_A^{\rm{exp}})/T_0$ can be computed for all transitions. The result is shown in Tab.~\ref{Tab:FitResults}. Based on these values we can renormalize the prefactors following $\nu_{\rm{des}}^{\rm{ren}} = \nu_{\rm des}e^{-\alpha/k_B}$. The resulting numbers calculated with $\nu_{\rm des} =  3\times 10^{13}$\,Hz are listed in Tab.~\ref{Tab:FitResults}. The good agreement with the experimental prefactors $\nu_{\rm{des}}^{\rm{exp}}$ shows that the model consistently describes all previous experimental studies. The large variation in frequencies and activation energies can be explained by assuming that the N flux affects the temperature dependence: it is largest if no N flux is present and monotonously decreases with increasing N flux (growth rate).

It is interesting to note that, although the temperature dependence has a huge effect on the experimentally measured apparent activation barrier $E_A^{\rm{exp}} = E_0 - \alpha T_0$ (by almost a factor of two) and prefactor $\nu_{\rm{des}}^{\rm{exp}}$ (by up to 12 orders of magnitude), the actual change in the effective desorption energy $E_A = E_0 + \alpha (T-T_0)$ is small: in the case of adsorption ($v = 0$), the activation barrier changes from 2.84\,eV to 2.73\,eV within the experimentally measured temperature window (700\,\deg C to 750\deg C). In the case of GaN growth, the variation is even smaller. This can be intuitively understood by considering that the experimentally measured activation energy represents a linear projection to zero temperature even when its real temperature dependence deviates strongly from linear behavior outside our experimental temperature window.

\section{Conclusion}

Based on a combination of experimental RHEED studies and first-principle growth models we have (i) quantitatively determined the Ga coverage on the GaN (0001) surface during adsorption as a function of Ga flux and substrate temperature and (ii) derived a model, which consistently describes the adsorption of Ga on GaN surfaces as well as the accumulation of Ga during Ga-rich GaN growth. This model resolves the discrepancy in previous measurements of the activation energy characterizing the critical Ga flux for the onset of Ga droplet fromation during GaN growth.\cite{Heying1,Adelmann}

The model also explains the origin of the experimentally-observed unphysically high prefactors in terms of a temperature dependent desorption barrier. At the moment, we can only speculate about
possible mechanisms which reduce the activation barrier at higher temperatures. A possible scenario emerges from our first principles calculations where we find that the number of atoms in the compressed Ga layer of the bilayer structure and thus the lateral lattice constant of the top Ga layer significantly changes with temperature. This change in the surface geometry is expected to have an important effect on the island formation energy and will be discussed in a forthcoming paper.\cite{LymperakisToBePublished}

\begin{acknowledgments}

The authors would like to thank O. Briot (University of Montpellier, France) for providing the GaN templates and F. Rieutord (CEA Grenoble, DRFMC/SI3M) for valuable discussions. E. Bellet-Amalric (CEA Grenoble, DRFMC/SP2M) is acknowledged for the analysis of Fig.~\ref{Fig:RHEED}. L. L. and J. N. like to thank the EU TMR program IPAM and J. N. the Deutsche Forschungsgemeinschaft SFB 296.

\end{acknowledgments}

\clearpage


\newpage

\begin{thebibliography}{99}

\bibitem{Jain} S. C. Jain, M. Willander, J. Narayan, and R. van Overstraeten, J. Appl. Phys. \textbf{87}, 965 (2000).

\bibitem{Hacke} P. Hacke, G. Feuillet, H. Okumura, and S. Yoshida, Appl. Phys. Lett. \textbf{69}, 2507 (1996).

\bibitem{Rapcewicz} K. Rapcewicz, M. Buongiorno Nardelli, and J. Bernholc, Phys. Rev. B \textbf{56}, R12725 (1997).

\bibitem{Zywietz} T. Zywietz, J. Neugebauer, and M. Scheffler, Appl. Phys. Lett. {\bf 73}, 487 (1998).

\bibitem{Smith} A. R. Smith, V. Ramachandran, R. M. Feenstra, D. W. Greve, M.-S. Shin and M. Skowronski, J. Neugebauer, and J. E. Northrup, J. Vac. Sci. Technol. B \textbf{16}, 1641 (1998); A. R. Smith, R. M. Feenstra, D. W. Greve, M. S. Shin, M. Skowronski, J. Neugebauer, and J. E. Northrup, \emph{ibid.} \textbf{16}, 2242 (1998).

\bibitem{Fritsch} J. Fritsch, O. F. Sankey, K. E. Schmidt, and J. B. Page, Phys. Rev. B \textbf{57}, 15360 (1998).

\bibitem{Xue} Q.-Z. Xue, Q.-K. Xue, R. Z. Bakhtizin, Y. Hasegawa, I. S. T. Tsong, T. Sakurai, and T. Ohno, Phys. Rev. B \textbf{59}, 12604 (1999); Q.-K. Xue, Q.-Z. Xue, R. Z. Bakhtizin, Y. Hasegawa, I. S. T. Tsong, T. Sakurai, and T. Ohno, Phys. Rev. Lett. \textbf{82}, 3074 (1999).

\bibitem{Munkholm} A. Munkholm, G. B. Stephenson, J. A. Eastman, C. Thompson, P. Fini, J. S. Speck, O. Auciello, P. H. Fuoss, and S. P. DenBaars, Phys. Rev. Lett. \textbf{83}, 741 (1999).

\bibitem{Northrup1} J. E. Northrup, J. Neugebauer, R. M. Feenstra, and A. R. Smith, Phys. Rev.  B \textbf{61}, 9932 (2000).

\bibitem{Wang} F.-H. Wang, P. Kr\"uger, and J. Pollmann, Phys. Rev. B \textbf{64}, 035305 (2001).

\bibitem{Tarsa} E. J. Tarsa, B. Heying, X. H. Wu, P. Fini, S. P. DenBaars, and J. S. Speck, J.  Appl. Phys. {\bf 82}, 5472 (1997).

\bibitem{Widmann} F. Widmann, B. Daudin, G. Feuillet, N. Pelekanos, and J. L. Rouvi\`ere, Appl. Phys. Lett. {\bf 73}, 2642 (1998).

\bibitem{Xie}  M. H. Xie, S. M. Seutter, W. K. Zhu, L. X. Zheng, H. Wu, and S. Y. Tong, Phys. Rev. Lett. \textbf{82}, 2749 (1999).

\bibitem{Heying1} B. Heying, R. Averbeck, L. F. Chen, E. Haus, H. Riechert, and J. S. Speck, J. Appl. Phys. {\bf 88}, 1855 (2000).

\bibitem{Bourret} A. Bourret, C. Adelmann, B. Daudin, J.-L. Rouvi\`ere, G. Feuillet, and G. Mula, Phys. Rev. B \textbf{63}, 245307 (2001).

\bibitem{Mula} G. Mula, C. Adelmann, S. Moehl, J. Oullier, and B. Daudin, Phys. Rev. B \textbf{64}, 195406 (2001).

\bibitem{Adelmann} C. Adelmann, J. Brault, D. Jalabert, P. Gentile, H. Mariette, G. Mula, and B. Daudin, J. Appl. Phys. {\bf 91}, 9638 (2002).

\bibitem{Heying2} B. Heying, I. Smorchkova, C. Poblenz, C. Elsass, P. Fini, S. DenBaars, U. Mishra, and J. S. Speck, Appl. Phys. Lett. {\bf 77}, 2885 (2000).

\bibitem{Kruse} C. Kruse, S. Einfeldt, T. B\"ottcher, D. Hommel, D. Rudloff, and J. Christen, Appl. Phys. Lett. {\bf 78}, 3827 (2001).

\bibitem{Zheng} L. X. Zheng, M. H. Xie, and S. Y. Tong, Phys. Rev. B \textbf{61}, 4890 (2001).

\bibitem{Brochard} F. Brochard-Wyart, J.-M. di Meglio, D. Qu\'er\'e, and P.-G. de Gennes, Langmuir \textbf{7}, 335 (1991).

\bibitem{footnote} Here, $\rho$ denotes the adatom density on top of the last completed Ga layer (a monolayer for the $1\rightarrow 2$ transition and a bilayer for the $2\rightarrow 3$ transition). This is contrasted by the Ga coverage $c$, which gives the \emph{total} amount of Ga present on the surface (including completed Ga mono- or bilayers).

\bibitem{StepsToBePublished} L. Lymperakis \emph{et al.} (unpublished).

\bibitem{Troullier} N. Troullier and J.L. Martins, Phys. Rev. B \textbf{43},
1993 (1991).

\bibitem{PBE} J.P. Perdew, K. Burke and M. Ernzerhof, Phys. Rev. Lett.
\textbf{77}, 3865 (1996); \textit{ibid.} \textbf{80}, 891 (1998).

\bibitem{Loui82} S.G. Louie, S. Froyen, M.L. Cohen, Phys. Rev. B \textbf{26},
1738 (1982).

\bibitem{Bockstedtexx} M. Bockstedte, A. Kley, J. Neugebauer and M. Scheffler,
Comput. Phys. Commun. \textbf{107}, 187 (1997).

\bibitem{Monkhorstxx} H.J. Monkhorst and J.D. Pack, Phys. Rev. B \textbf{13},
5188 (1976).

\bibitem{mrsProc} C. D. Lee, R. M. Feenstra, J. E. Northrup, L. Lymperakis, J. Neugebauer, Mater. Res. Soc. Proc. \textbf{743}, L4.1 (2002)

\bibitem{mPlane} C. D. Lee, R. M. Feenstra, J. E. Northrup, L. Lymperakis, J. Neugebauer, Appl. Phys. Lett. (in print).

\bibitem{Fuchs}M. Fuchs, J. L. F. Da Silva, C. Stampfl, J. Neugebauer, M.
Scheffler, Phys. Rev. B \textbf{65}, 245212 (2002).

\bibitem{spinPolarization} The spin-polarized correction given by \\ \texttt{http://math.nist.gov/DFTdata/atomdata/tables/ptable.html} \\ has been used.

\bibitem{Wyckoff} R. W. G. Wyckoff, \textit{Crystal Structures} (Wiley, New York, 1962), Vol. 1, 2$^\mathrm{nd}$ ed.

\bibitem{Kittel} C. Kittel, \textit{Introduction to Solid State Physics} (Wiley, New York, 1986).

\bibitem{LymperakisToBePublished} L. Lymperakis \emph{et al.} (unpublished).

\end{thebibliography}
\end{document}